# The Impact of Dormant Defects on Defect Prediction: a Study of 19 ApacheProjects

**DAVIDE FALESSI,** University of Rome Tor Vergata, Italy

**AALOK AHLUWALIA,** California Polytechnic State University, USA

**MASSIMILIANO DI PENTA,** University of Sannio, Italy

Defect prediction models can be beneficial to prioritize testing, analysis, or code review activities, and has been the subject of a substantial effort in academia, and some applications in industrial contexts. A necessary precondition when creating a defect prediction model is the availability of defect data from the history of projects. If this data is noisy, the resulting defect prediction model could result to be unreliable. One of the causes of noise for defect datasets is the presence of "dormant defects", i.e., of defects discovered several releases after their introduction. This can cause a class to be labeled as defect-free while it is not, and is, therefore "snoring". In this paper, we investigate the impact of snoring on classifiers' accuracy and the effectiveness of a possible countermeasure, i.e., dropping too recent data from a training set. We analyze the accuracy of 15 machine learning defect prediction classifiers, on data from more than 4,000 defects and 600 releases of 19 open source projects from the Apache ecosystem. Our results show that on average across projects: (i) the presence of dormant defects decreases the recall of defect prediction classifiers, and (ii) removing from the training set the classes that in the last release are labeled as not defective significantly improves the accuracy of the classifiers. In summary, this paper provides insights on how to create defects datasets by mitigating the negative effect of dormant defects on defect prediction.

## 1 INTRODUCTION

Defect prediction models aim at identifying software artifacts that are likely to exhibit a defect [47, 54–56, 72, 75]. The main purpose of defect prediction is to reduce the cost of testing, analysis, or code review, by prioritizing developers' effort on specific artifacts. In this area, researchers have proposed the use of different models to predict defects, leveraging for example product metrics [4, 24, 26, 34, 52], process metrics [48], the knowledge of where previous defects occurred [37, 56], information about change-inducing fixes [33, 35] and, recently, using deep learning techniques to automatically engineer features from source code elements [74].

Authors' addresses: Davide Falessi, falessi@ing.uniroma2.it, University of Rome Tor Vergata, Italy; Aalok Ahluwalia, ahluwali@calpoly.edu, California Polytechnic State University, USA; Massimiliano Di Penta, dipenta@unisannio.it, University of Sannio, Italy.

Several researchers have worked on improving the performance of defect estimation models using techniques such as tuning [20, 23, 70], re-balancing [1, 5], or feature selection [71]. In order to promote the usage and improvement of prediction models, researchers have provided means to create [21, 81], collect [16] and select [22, 50, 63] defect datasets. Ultimately, the reliability of a prediction model depends on the quality of the dataset [40, 65]. Previous work has identified sources of noise in datasets, including defect misclassification [6, 27, 36, 60, 69] and defect origin, [62] and proposed solutions to deal with them.

In summary, while researchers are attempting to build better models for defect prediction, the quality of the datasets used to train such a model plays a paramount role in obtaining accurate and usable results. To predict which classes are defect-prone, defect prediction classifiers need to be trained on a dataset where classes in releases have been previously labeled as defective or not. The best way to identify when a defect has been introduced in a software project is to use a "realistic approach", i.e., using the affected version reported in Jira [79]. Another viable approach is the SZZ algorithm [38, 66]. SZZ exploits the versioning system annotation mechanism (e.g., *git blame*) to determine, for the source code lines that have been changed in a defect fix, when they have last been changed before such a fix. On the one hand, assigning defects onto releases based on Jira fully reflects the outcome of the bug triaging, however such information, in existing datasets, not only is fairly incomplete, but, also, it may not be available immediately as soon as a bug is discovered. On the other hand, SZZ is particularly useful in complementing the realistic approach, i.e., SZZ can be used for defects where the affect version is not reported in Jira. At the same time, it is known to be subject to imprecisions [13].

Regardless of the specific approach to identify the origin of defects, defects need to be fixed before any approach can be used to label classes as defective or not. Specifically, a defect may be discovered several releases after its introduction; this phenomenon has been called "dormant defect" [8, 61]. Thus, if we observe today the status of a class in its current version, it can be labeled as not defective while it is actually defective. Thus, a class has a noise that we call "snoring" under two conditions: (i) it contains at least one dormant defect, and (ii) it is labeled while all contained defects are still dormant.

In a recent paper [3] we found that most of the defects in a project slept for more than 20% of releases and that the proportion of defective classes that are not identified as defective is more than 25% even if we ignore 50% of releases. Now that we know that defects sleep and datasets snore, we want to know and inhibit their negative impact on defect prediction. Thus, in this study, we address the following research questions:

- **RQ1:** *Does snoring reduce defect prediction performance?* We conjecture that the presence of snoring classes can bias classifiers during training and hence can negatively impact the classifiers' performance. In this research question, we investigate the extent to which the snoring that exists in open source projects impacts the performance of classifiers in predicting defect-prone classes.
- **RQ2:** *Does defect prediction performance improve if we remove from the training set the classes that are labeled as non-defective in the last releases?* Snoring classes are, by definition, more frequent in more recent releases, i.e., in releases closer to when classes are labeled. Moreover, classes labeled as defective cannot snore. This happens

  because such recent releases contain defects that may not have been discovered yet. Therefore, we might remove noise and hence improves the performance of the classifiers by ignoring in the training set the classes that in the last releases are labeled as not defective. At the same time, if we reduce too much the training set by removing several recent releases because they are potentially affected by snoring, the limited availability of data would also affect the classifiers' performance. In other words, is it better to train a classifier on a big noisy dataset or on a less noisy yet smaller dataset? Where is the sweet spot? In this research question, we measure the impact of a countermeasure for snoring, which consists of removing from the training set the classes that in the last releases are labeled as not defective.

In our empirical study, we analyze data from 19 Apache projects featuring a total of more than 4,000 defects and 600 releases. Also, we compare the performance of 15 different machine learning classifiers in the context of within-project across-release defect prediction. Our results show that:

(1) Snoring decreases the performance of each of the 15 classifiers. More specifically, while precision is not sig- nificantly affected, the presence of snoring significantly affects recall, with a large effect size on each of the 15 classifiers. The effect size of snoring on other performance indicators, such as the Matthews Correlation Coefficient (MCC), the Area Under the Receiving Operating Characteristics Curve (AUC), and Cohen Kappa is medium or large.

(2) Removing from the training set the classes that are labeled as non-defective in the last release increases the classifiers' performance. For instance, Recall, F1, Kappa, and Matthews increase by about 30%. Removing a further release could further improve performance, however going behind worsen the performances, because the classifiers are trained with less and outdated data.

The remainder of the paper is structured as follows. Section 2 describes the empirical study design. Section 3 reports and discusses the results of the empirical investigation. Section 5 discusses the threats to the study validity. Section 6 discusses the related literature, focusing in particular on noise and imprecision on defect prediction, and on techniques to build defect prediction datasets. Finally, Section 7 concludes the paper and outlines directions for future work.

## 2 EMPIRICAL STUDY DESIGN

The *goal* of this study is to investigate the effect of snoring data on classifiers' performance, their evaluation, and the effectiveness of a countermeasure. The *quality focus* is the performance of such models, which can be affected by snoring. The *perspective* of the study is of researchers or practitioners that want to improve the way defect prediction datasets are constructed so that classifiers can more accurately predict defect-prone classes.

### 2.1 Definition of Snoring Classes

To better understand when and how classes snore, let us consider a project with three releases, $r1$, $r2$, and $r3$. If a defect has been introduced in $r1$ and only fixed in $r3$, then the presence of the defect would not be considered in $r1$ and $r2$. If in $r1$ and $r2$ a class does not exhibit any other defect but the dormant defect in question, such a class will be erroneously labeled as non-defective in $r1$ and $r2$. This class is a false negative and it is a snoring noise in defect dataset. If in $r1$ and $r2$ a class does exhibit at least another defect, then the defect sleeping in $r1$ and $r2$ does not impact the labeling of the class since this class is labeled as defective anyway. This class is a true positive. Table 1 reports a scenario of a project in which events related to defects (I = introduced, N = nothing, and F = fixed) happen in three releases (columns) and impact three classes (rows). A defect is defined as a post-release defect if it is fixed in a release after the one it has been introduced. Thus, a defective class is a class having at least one post-release defect. For instance, in column 2, row 2 of Table 1, a defect in class C1 is introduced and fixed in the same release, $r1$. Thus, C1 is not defective as it contains no post-release defect.

Based on the events described in Table 1, Table 2 reports the (post-release) defectiveness for a class, when it is labeled at $r2$ (left-side), or $r3$ (right-side). Each class in each release is labeled as defective (D) or not defective (ND). It is

Table 1. An example of a project in which events (I = introduced, N = nothing, and F = fixed) happen in three releases and three different classes.

|    | r1 | r2 | r3 |
|----|----|----|----|
| C1 | IF | I  | F  |
| C2 | I  | N  | F  |
| C3 | II | F  | F  |

Table 2. Post-release defectiveness of a class in a specific release.

Dataset created at *r2*

| Classes | r1 |
|---------|----|
| C1      | ND |
| C2      | ND |
| C3      | D  |

Dataset created at *r3*.

| Classes | r1 | r2 |
|---------|----|----|
| C1      | ND | D  |
| C2      | D  | D  |
| C3      | D  | D  |

important to note that the label of C2 in *r1* changes on whether it is labeled at the end of *r2* or *r3*. Specifically, if C2 is labeled at the end of *r2*, then the introduction is not discovered and hence the class C2 at *r1* is labeled as not defective.

Table 3. How dormant defect bias datasets

Dataset created at *r2*.

| Classes | r1 |
|---------|----|
| C1      | TN |
| C2      | FN |
| C3      | TP |

Dataset created at *r3*.

| Classes | r1 | r2 |
|---------|----|----|
| C1      | TN | TP |
| C2      | TP | TP |
| C3      | TP | TP |

Table 3 reports the snoring caused by snoring, of a class in a release, according to when the dataset is created, in terms of FN = the class is erroneously marked as not defective despite being defective, TN = the class is marked as not defective and is not defective and TP = the class is marked as defective. Specifically, C2 at *r1* is a FN for a dataset created at the end of *r2* (left-side table) and a TP for dataset created at *r3* (right-side table). Note that we consider FN and do not consider FP because the dormant defects cannot introduce FP, i.e., if a class is labeled as defective then one or more defects have been fixed already and no defect that is possibly discovered in the future can change the class label to non-defective. Finally, we note that the discussion above applies to all known labeling mechanisms such as SZZ [38, 66] and the realistic one [79]. The presence of snoring in the data is due to when the classes are labeled as defective or not, and it is entirely independent of the specific technique used to label classes such as the realistic or SZZ. As a matter of fact, both realistic and SZZ use the fixed tickets as input; thus, both approaches entirely fail in taking into consideration the bugs that are not fixed yet.

The study *context* consists of data from 19 open source projects from the Apache ecosystem. We focused on Apache[1] projects rather than random GitHub projects because the former have a higher quality of defect annotation, and this also avoided us to consider unrealistic/toy projects [49]. We select the 19 projects that are managed in Jira, versioned in Git, have at least ten releases, have most of the commits related to Java code, and have the highest proportion of defects linked to commits in the source code. A defect is linked if it can be associated with at least one commit in the source code's commit log. Table 4 reports the details of the used 19 projects in terms of releases, days, commits, defects, and proportion of defects providing the affected version in Jira.

---
[1] https://people.apache.org/phonebook.html

Table 4. Details of the used projects.

| Project | Releases | Days | Commits | Defects | AV in Jira |
|---|---|---|---|---|---|
| AVRO | 46 | 3528 | 1770 | 114 | 0.60 |
| BOOKKEEPER | 22 | 2880 | 2056 | 184 | 0.32 |
| CHUKWA | 11 | 3638 | 849 | 7 | 0.47 |
| CONNECTORS | 118 | 3311 | 4672 | 261 | 0.85 |
| CRUNCH | 17 | 2663 | 1055 | 132 | 0.48 |
| FALCON | 31 | 2450 | 2227 | 219 | 0.40 |
| GIRAPH | 10 | 2719 | 1096 | 123 | 0.29 |
| IVY | 17 | 4977 | 2973 | 133 | 0.89 |
| OPENJPA | 31 | 4671 | 4978 | 380 | 0.82 |
| PROTON | 51 | 2513 | 3929 | 53 | 0.69 |
| SSHD | 26 | 3652 | 1589 | 124 | 0.64 |
| STORM | 36 | 2634 | 9754 | 442 | 0.52 |
| SYNCOPE | 48 | 3214 | 6320 | 296 | 0.93 |
| TAJO | 13 | 2407 | 2273 | 286 | 0.23 |
| TEZ | 34 | 2163 | 2661 | 559 | 0.14 |
| TOMEE | 22 | 4792 | 12135 | 277 | 0.27 |
| WHIRR | 8 | 1788 | 569 | 20 | 0.37 |
| ZEPPELIN | 14 | 2067 | 4048 | 200 | 0.37 |
| ZOOKEEPER | 49 | 3923 | 1820 | 219 | 0.71 |

In the following, we describe the empirical design of each research question.

## 2.2 RQ1: Does snoring reduce defect prediction performance?

We are in the shoes of a practitioner who uses previous project releases' data to predict which classes of the current release are defect-prone. Thus, we want to measure the performance of the same classifier when trained on a dataset with versus without snoring. In this research question, we test the null hypothesis: $H_{01}$: there is no difference between the prediction performance achieved by classifiers when trained on a dataset with snoring versus a dataset without snoring.

### 2.2.1 Independent Variables.
The independent variable of this research question is the presence or absence of snoring in the training set.

### 2.2.2 Dependent Variables.
The dependent variable is the relative loss (RL) that we measure as the relative distance between the performance of a classifier trained on a dataset with snoring versus trained on a dataset without snoring. Specifically,

$$RL = |performance_{with\ snoring} - performance_{without\ snoring}| / performance_{without\ snoring}.$$

Thus, the relative loss represents the detriment on performance caused by snoring, as observed by each different performance metric. As performance metrics we used the following metrics:

- True Positive (TP): The class is actually defective and is predicted to be defective.

- False Negative (FN): The class is actually defective and is predicted to be non-defective.
- True Negative (TN): The class is actually non-defective and is predicted to be non-defective.
- False Positive (FP): The class is actually non-defective and is predicted to be defective.
- **Precision**: $\frac{TP}{TP+FP}$
- **Recall**: $\frac{TP}{TP+FN}$
- $F_1$: $\frac{2 \cdot Precision \cdot Recall}{Precision + Recall}$
- Cohen's **Kappa**: A statistic that assesses the classifier's performance against random guessing [10]. $Kappa = \frac{Observed - Expected}{1 - Expected}$
  - Observed: The proportionate agreement. $\frac{TP+TN}{TP+TN+FP+FN}$
  - Expected: The probability of random agreement. $PP_{Yes} + PP_{No}$ where
    * $PP_{Yes}$: Probability of positive agreement. $\frac{TP+FP}{TP+TN+FP+FN} \cdot \frac{TP+FN}{TP+TN+FP+FN}$
    * $PP_{No}$: Probability of negative agreement. $\frac{TN+FP}{TP+TN+FP+FN} \cdot \frac{TN+FN}{TP+TN+FP+FN}$
- **AUC** (Area Under the Receiving Operating Characteristic Curve) [58] is the area under the curve, of true-positives rate versus false positive rate, that is defined by setting multiple thresholds. AUC has the advantage to be threshold independent.
- **MCC** (Matthews Correlation Coefficient) is commonly used in assessing the performance of classifiers dealing with unbalanced data [44], and is defined as: $\sqrt{\frac{TP \cdot TN - FP \cdot FN}{(TP+FP)(TP+FN)(TN+FP)(TN+FN)}}$. Its interpretation is similar to correlation measures, i.e., $MCC < 0.2$ is considered to be low, $0.2 \leq MCC < 0.4$—fair, $0.4 \leq MCC < 0.6$—moderate, $0.6 \leq MCC < 0.8$—strong, and $MCC \geq 0.8$—very strong.

2.2.3 *Measurement.* Our measurement procedure consists of the following steps:

(1) *Labeling defective classes*: For each project, we created a dataset **D** by labeling classes in releases as defective or not. To do so, we used the "realistic approach" [79], i.e., using the affected version reported in Jira. This means that, given our point of observation (which, as explained below and shown as a photo camera in Fig. 2, is where a training set is built), we consider the affected version from all bug reports opened at that time. for each bug, we consider as affected versions those indicated by in the bug report. Moreover, since the realistic approach is based on the use of the affected version that is reported in Jira, and since according to Table 4 a considerable number of defects do not report the affected version in Jira, then, using only the realistic approach would have produced several false negative labels of classes. Therefore, to avoid noise in our datasets due to missing affected versions in JIRA, we used SZZ for defects where the affected version is not available. We re-implemented the SZZ algorithm [66] and improved it by ignoring comments[2], indentation, white spaces, and documentation strings[3] as changes introducing defects. Thus, similarly to what we have done in a recent paper [73], we identify the introduction version of each defect using SZZ if the ticket was without affected version, or as the oldest of the affected versions in the ticket. Afterward, regardless of how we identified the introduction version, we label the classes involved in the fix commit of the ticket as defective in releases in the range [introduction version, fixed version).

---
[2] https://goo.gl/X8fHFc
[3] https://goo.gl/dNXb6N

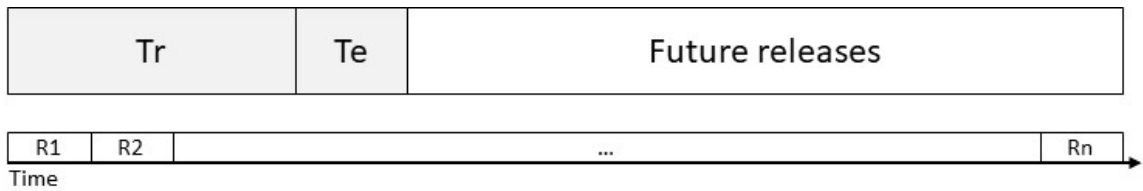

Fig. 1. The evaluation technique adopted in this paper. The first 66% of data is used as training set, the remaining 33% as test set. Several releases are unused to avoid that snoring impacts our results.

(2) *Avoiding snoring in our measurement*: Since in a recent paper [3] we proved that removing most recent releases implies removing snoring, then, for each project, we created SD by removing the last 50% of releases from D. Therefore, by using SD, rather than D, we are confident that our ground truth is not significantly affected by snoring.

(3) *Training and test sets*: Fig. 1 describes our overall validation technique which is then customized for the tworesearch questions. Our overall validation technique is a 66/33 holdout that preserves the order of data [15]. First,the data is ordered according to the releases of the project, i.e., we created D. Second, the last 50% of releasesare removed to avoid that snoring impacts our ground truth by more than 1% [3], i.e., we created SD. Third, thefirst 66% of SD is used as a training set (Tr), the remaining 33% of SD is used as a test set (Te). Fig. 2 reportsthe validation technique used in this research question. In the top part of Fig. 2 we have the ideal scenario: wecreate the test sets and training sets without snoring (i.e., TrNS and TeNS) by executing the labeling technique,which is represented with a photo camera, at the end of the project. On the bottom part of Fig. 2 we replicatethe realistic scenario: we create the training sets with snoring (i.e., TrS) by executing the labeling technique,which is represented with a photo camera, at data collection, i.e., at the end of the training set. For example, inproject Bookkeeper, Tr is constructed using the first eight releases. The class BookieProtocol.java in releases seven and eight of TrS is marked as non-defective, but will become defective if observed after release 11 because of the defect BOOKKEEPER-1018[4]. Thus, the same exact class BookieProtocol.java in releases seven and eight islabeled as defective in TrNS and as non-defective in TrS.

(4) *Classifiers' performance*: We train each classifier on TrS vs. TrNS and tested on TeNS. Thus, we compare the performance of the same classifier when trained with snoring versus trained and without snoring.

We use 16 well-defined product and project metrics that previous work found to be useful for defect prediction [14, 17]. These predictor metrics are detailed in Table 5. We note that TrS and TrNS share all predictor metrics and values, and differ only in the defectiveness label of some classes, since classes are labeled at different pointsof time, in the training set. Afterward, we perform four sub-steps.

(a) Feature Selection. We filter the predictor variables described above by using the correlation-based feature subset selection [25]. The approach evaluates the worth of a subset of attributes by considering the individual predictive capability of each feature, as well as the degree of redundancy between different features. The approach searches the space of attribute subsets by greedy hill-climbing augmented with a backtracking facility. The approach starts with the empty set of attributes and performs a forward search by considering all possible single attribute additions and deletions at a given point.

---
[4] https://issues.apache.org/jira/browse/BOOKKEEPER-1018

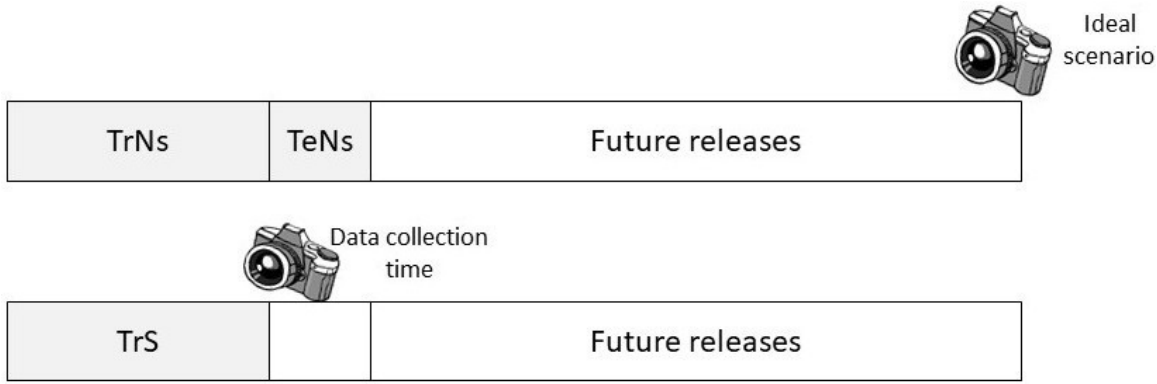

Fig. 2. A photo camera represents the technique that is used to label class defectiveness. We executed the technique at different points in time. On the bottom side, the labeling is performed in a realistic usage scenario, i.e., when one realistically collects the data to be used in the prediction model. This execution produces a training set with snoring (TrS). On the top side, the collection is performed in an ideal scenario, i.e., at the end of the project. This execution produces a training and test set without snoring (TrNSand TeNS).

Table 5. Defect prediction features.

| Metric | Description |
|---|---|
| Size | Lines of code(LOC) |
| LOC Touched | Sum over revisions of LOC added + deleted |
| NR | Number of revisions |
| Nfix | Number of bug fixes |
| Nauth | Number of authors |
| LOC Added | Sum over revisions of LOC added + deleted |
| MAX LOC Added | Maximum over revisions of LOC added |
| AVG LOC Added | Average LOC added per revision |
| Churn | Sum over revisions of added – deleted LOC |
| Max Churn | Maximum churn over revisions |
| Average Churn | Average churn over revisions |
| Change Set Size | Number of files committed together |
| Max Change Set | Maximum change set size over revisions |
| Average Change Set | Average change set size over revisions |
| Age | Age of Release |
| Weighted Age | Age of Release weighted by LOC touched |

(b) Classifiers. As classifiers we use the 14 used in a previous related paper [32]:

- Decision Stump: A single level decision tree performing classification based on entropy [30].
- Decision Table: Two major parts: schema, the set of features included in the table, and a body, labeledinstances defined by features in the schema. Given an unlabeled instance, try matching instance to record in
  the table. [41]
- IBk: K-nearest neighbors classifier run with k = 1 [2].
- J48: Generates a pruned C4.5 decision tree [59].

- JRip: A propositional rule learner, Repeated Incremental Pruning to Produce Error Reduction (RIPPER) [11].
- KStar: Instance-based classifier using some similarity function. Uses an entropy-based distance function [9].
- Naive Bayes: Classifies records using estimator classes and applying Bayes theorem [31]
- Naive Bayes Updateable: An instance of the Naive Bayes classifier with different weight initial values and constraints [31]
- OneR: 1R classifier using the minimum-error attribute for prediction [29].
- PART: Uses separate-and-conquer, building partial C4.5 decision trees and turning the best leaf into a rule [19].
- Random Forest: Ensemble learning creating a collection of decision trees. Random trees correct for overfitting [7].
- REPTree: Fast decision tree learner. Builds a decision tree using information gain and variance, and prunes using reduced-error pruning [43].
- SMO: John Platt's sequential minimal optimization algorithm for training a support vector classifier [57]

(c) Tuning. Instead of tuning each single classifier, we used AutoWEKA [42], an automated approach for classifier selection and hyperparameter optimization. We run AutoWEKA for two hours on each training set (with or without snoring) of each project. Thus we refer to *AutoWEKA* as the classifier and parameter selected by AutoWEKA for the specific training set.

(d) Test. We train each of the 14 classifiers, plus the AutoWEKA classifier, on TrS and test on TeNS and we train on TrNS, and we test on TeNS.

It is possible that snoring could depend on the distance between releases, measured in terms of days or commits. To verify whether $RQ_1$ results are affected by that, we compute the Spearman's rank correlation [67] between the number of days and commits per release and the model's relative loss. A high and statistically significant, positive or negative correlation, could indicate that results may depend on the distance between releases, i.e., snoring is more likely to occur for rapid releases (and therefore one may need to drop more releases) than for a longer distance between releases.

*2.2.4 Analysis Procedure.* We compare the accuracy of models when trained on a training set with vs. without snoring (i.e., TrS vs. TrNS) and tested on the test set without snoring (TeNS).

Since our data strongly deviate from normality, the hypotheses of this research question are tested using the Wilcoxon signed-rank test [76], and Cliff's delta (paired) effect size [23]. Also, since we repeat the statistical analysis multiple times (i.e., for each classifier), we adjust p-values using the Holm's correction procedure [28]. Such a procedure ranks $oo$ p-values in increasing order of value and multiplies the first one (i.e., the smaller) by $oo$, the second by $oo - 1$, and so on.

## 2.3 RQ$_2$: Does defect prediction performance improve if we remove from the training set the classes that are labeled as non-defective in the last releases?

We are in the shoes of a practitioner who reflects on whether training classifiers on less but non-noisy data improves the defect prediction performance over training classifiers on all data. At the time of the prediction the practitioner does not know which specific class snores; the only information they can leverage to remove snoring classes is that i) younger releases are more likely to contain snoring classes than older releases, and 2) classes labeled as not defective cannot be snoring. Thus, the only feasible and realistic approach that can be used to remove snoring, at the time of the prediction, is to remove all the classes labeled as defective in the last releases. However, this approach, since it is not perfectly accurate, is a double-edged sword; on the one side it removes all snoring data in the last releases, on the other side it removes data that are also not snoring (i.e., actually defective classes). To address this research question, we test the null hypothesis: $H_{02}$: there is no difference between the prediction performance achieved by classifiers when trained on an entire dataset versus a portion of the same dataset where classes labeled as non-defective are removed from last releases.

*2.3.1 Independent Variables.* The independent variable of $RQ_2$ is the number of releases from which classes labeled as non-defective are removed from the training set.

*2.3.2 Dependent Variables.* The dependent variable is the relative loss of the performance measures used to address $RQ_2$.

*2.3.3 Measurement Procedure.* Fig. 3 reports the technique used to replicate the scenario where the user wants to use a defect prediction model and they tries to improve the prediction accuracy by removing from the training set the classes labeled as not defective in last releases. Thus, as in RQ1, the technique used to label class defectiveness is executed at the point in time where the model is used for prediction, i.e., at the end of the training set. Differently from RQ1, we remove from the training set the classes labeled as not defective in the last 0,1,2,3, and 4 releases; we call these training sets TrS, TrS-1, TrS-2, TrS-3, and TrS-4 respectively. As in RQ1, the training and test sets without snoring (TrNS and TeNS) are created by executing the labeling technique at the end of the project.

For example, removing classes labeled as not defective from the last release of the training set of the *Bookkeeper* project, to form TrS-1, would eliminate classes that snore at that release such as `ZkLedgerUnderreplicationManager.java` [5].

*2.3.4 Analysis Procedure.* We compare the accuracy of models when trained on training sets obtained by removing all classes labeled as not defective from the last 0,1,2,3, and 4 releases and tested on a test set without snoring (TeNS).

Then, we test whether there is a correlation between the removal of non-defective classes of the last release(s) of the training set and the gain in terms of Precision, Recall, Cohen's kappa, MCC, and AUC. To this aim, we use a repeated-measure permutation test, using the Project as a random variable, and considering as an independent variable the number of removed releases and the choice of the classifier.

We also analyze the correlation between the gain in performance, provided by removing non-defective classes in the last release of the training set, and the frequency of releases in projects.

Finally, to promote replicability, we made available online all datasets, SZZ, and other developed scripts [6].

## 3 EMPIRICAL STUDY RESULTS

In the following, we report the results of our study to address the research questions formulated in Section 1.

### 3.1 RQ$_1$: Does snoring reduce defect prediction performance?

Table 6 reports, for each project, the number of classes that are labeled as Defective and Not Defective in TrNs, as Defective in TrS, and the loss as computed as the proportion of defective classes that are erroneously not labeled as defective in the TrS, i.e., (number of defective classes in TrNS - number of defective classes in TrS) / number of defective classes in TrNS). According to Table 6, for most of the projects, the majority of defective classes are affected by snoring (Loss > 50%). In the remainder of this research question, we observe how this loss affects defect prediction models.

---
[5] https://tinyurl.com/ybupvq3q
[6] http://doi.org/10.5281/zenodo.4394035

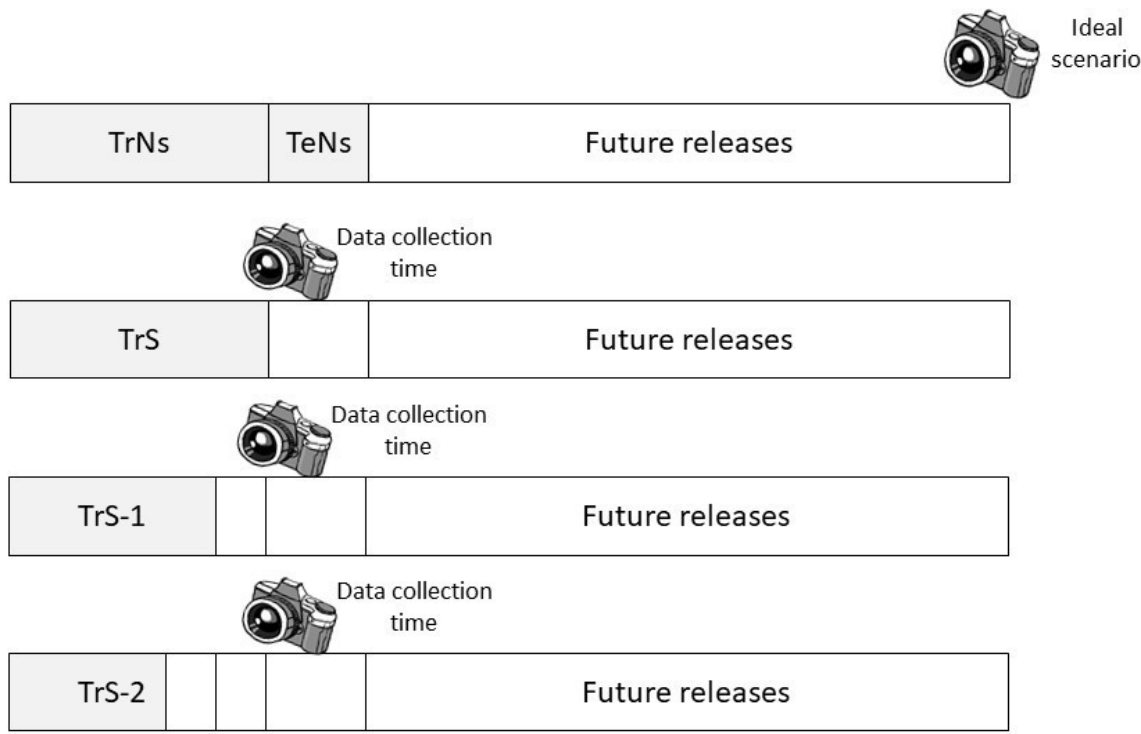

Fig. 3. The technique used to replicate the scenario where the user wants to use defect prediction model and tries to improve its accuracy by removing classes labeled as not defective from the last 0,1,2,3, and 4 releases of the training set with snoring.

Figure 4 reports, as boxplots, the distribution of Precision, Recall, F1-Score, AUC, MCC, and Cohen $kk$ for different classifiers and projects, achieved when trained on data with (red boxes) versus without (blue boxes) snoring. This same data is reported for each classifier in Figure 5. According to Figure 5 the presence of snoring in the training set decreases the performance in each of the 15 classifiers and in each of the six performance metrics. Figure 6 reports, as boxplots, the distribution of relative loss in Precision, Recall, AUC, F1-Score, Cohen $kk$ and MCC for different classifiers and projects. According to Figure 6, the highest median relative loss, among performance metrics, is in Recall, and it is about 90%.

As explained in Section 2, the observed differences have been statistically analyzed using Wilcoxon signed-rank and Cliff's delta effect size (positive effect sizes indicate a difference in favor of results without snoring). As Table 7 shows, Recall is significantly higher without snoring, as confirmed in Figure 4, with an effect size that is always large. Instead, we found no significant differences in what concerns Precision.

Table 7 and Table 8 report the statistical test results about the differences in performances of defect prediction classifiers when trained on a dataset with versus without snoring. As Table 7 and Table 8 show, significant differences in terms of all performance metrics other than Precision are observed for all classifiers other than decision_stump. Thus, we can reject $H_10$ and claim that **snoring significantly worsens the accuracy of all classifiers and all metrics other than Precision**.

Table 6. Snoring classes: proportion of defective classes not labeled as defective in the training set.

| Project | TrNs | | TrS | Loss |
|---|---|---|---|---|
| | Defective | Not Defective | Defective | |
| AVRO | 54 | 1249 | 17 | 31% |
| BOOKKEEPER | 836 | 2924 | 167 | 20% |
| CHUKWA | 1031 | 1681 | 1010 | 98% |
| CONNECTORS | 784 | 3963 | 690 | 88% |
| CRUNCH | 298 | 1460 | 152 | 51% |
| FALCON | 199 | 368 | 193 | 97% |
| GIRAPH | 592 | 2991 | 154 | 26% |
| IVY | 603 | 4517 | 271 | 45% |
| OPENJPA | 1000 | 12108 | 640 | 64% |
| PROTON | 350 | 10480 | 140 | 40% |
| SSHD | 346 | 991 | 242 | 70% |
| STORM | 2208 | 5220 | 1700 | 77% |
| SYNCOPE | 11202 | 9170 | 2801 | 25% |
| TAJO | 1261 | 4660 | 227 | 18% |
| TEZ | 486 | 914 | 408 | 84% |
| TOMEE | 326 | 5914 | 293 | 90% |
| WHIRR | 45 | 810 | 17 | 38% |
| ZEPPELIN | 786 | 2253 | 424 | 54% |
| ZOOKEEPER | 434 | 1693 | 308 | 71% |

Table 7. Impact of snoring on classifiers' performance ($RQ_1$): Wilcoxon signed-rank test results and Cliff's delta effect size on precision and recall for different machine learning classifiers.

| Classifier | Precision | | | Recall | | |
|---|---|---|---|---|---|---|
| | $p$-value | $o$ | Magnitude | $p$-value | $o$ | Magnitude |
| autoweka | 1.00 | 0.11 | negligible | 0.00 | 0.76 | large |
| decision_stump | 0.18 | 0.42 | medium | 0.01 | 0.58 | large |
| decision_table | 0.34 | 0.36 | medium | 0.01 | 0.64 | large |
| ibk | 0.07 | 0.38 | medium | 0.00 | 0.87 | large |
| j48 | 0.34 | 0.30 | small | 0.01 | 0.60 | large |
| jrip | 1.00 | 0.22 | small | 0.00 | 0.76 | large |
| kstar | 1.00 | 0.21 | small | 0.01 | 0.58 | large |
| naive_bayes | 1.00 | -0.05 | negligible | 0.00 | 0.65 | large |
| naive_bayes_update | 1.00 | -0.05 | negligible | 0.00 | 0.65 | large |
| oner | 0.29 | 0.42 | medium | 0.00 | 0.79 | large |
| part | 1.00 | 0.17 | small | 0.00 | 0.64 | large |
| random_forest | 0.04 | 0.34 | medium | 0.00 | 0.90 | large |
| random_tree | 0.34 | 0.15 | small | 0.00 | 0.87 | large |
| reptree | 0.02 | 0.48 | large | 0.00 | 0.93 | large |
| smo | 0.01 | 0.55 | large | 0.00 | 0.76 | large |

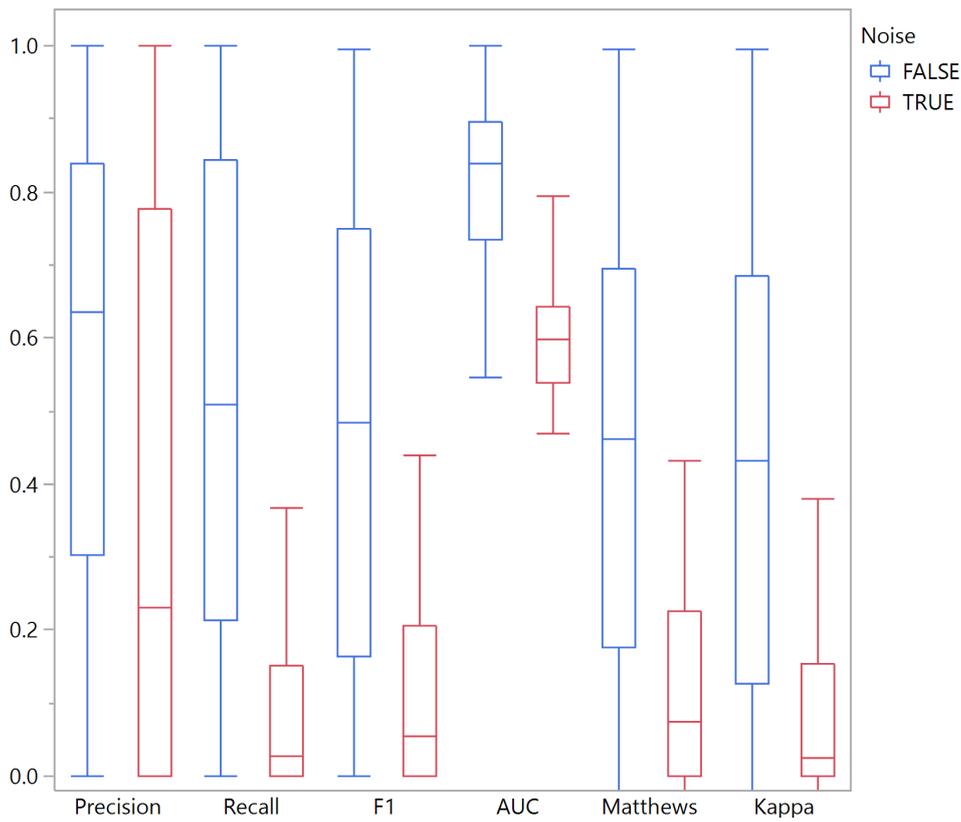

Fig. 4. Distribution of performance, among classifiers and datasets, achieved with versus without snoring.

Table 8. Impact of snoring on classifiers' performance (RQ$_1$): Wilcoxon signed-rank test results and Cliff's delta effect size on Cohen $kk$, MCC and AUC for different machine learning classifiers.

| Classifier | Cohen $kk$ | | | MCC | | | AUC | | |
|---|---|---|---|---|---|---|---|---|---|
| | $p$-value | $\delta$ | Magnitude | $p$-value | $\delta$ | Magnitude | $p$-value | $\delta$ | Magnitude |
| autoweka | 0.00 | 0.62 | large | 0.00 | 0.64 | large | 0.00 | 0.84 | large |
| decision_stump | 0.02 | 0.45 | medium | 0.03 | 0.45 | medium | 0.00 | 0.87 | large |
| decision_table | 0.00 | 0.53 | large | 0.00 | 0.54 | large | 0.00 | 0.84 | large |
| ibk | 0.00 | 0.80 | large | 0.00 | 0.78 | large | 0.00 | 0.82 | large |
| j48 | 0.01 | 0.52 | large | 0.02 | 0.51 | large | 0.00 | 0.82 | large |
| jrip | 0.02 | 0.60 | large | 0.03 | 0.58 | large | 0.00 | 0.82 | large |
| kstar | 0.00 | 0.58 | large | 0.00 | 0.54 | large | 0.00 | 0.82 | large |
| naive_bayes | 0.00 | 0.57 | large | 0.00 | 0.48 | large | 0.00 | 0.85 | large |
| naive_bayes_update | 0.00 | 0.57 | large | 0.00 | 0.48 | large | 0.00 | 0.84 | large |
| oner | 0.00 | 0.67 | large | 0.00 | 0.67 | large | 0.00 | 0.83 | large |
| part | 0.02 | 0.48 | large | 0.03 | 0.48 | large | 0.00 | 0.84 | large |
| random_forest | 0.00 | 0.80 | large | 0.00 | 0.78 | large | 0.00 | 0.84 | large |
| random_tree | 0.00 | 0.71 | large | 0.00 | 0.71 | large | 0.00 | 0.84 | large |
| reptree | 0.00 | 0.83 | large | 0.00 | 0.82 | large | 0.00 | 0.85 | large |
| smo | 0.00 | 0.75 | large | 0.00 | 0.75 | large | 0.00 | 0.84 | large |

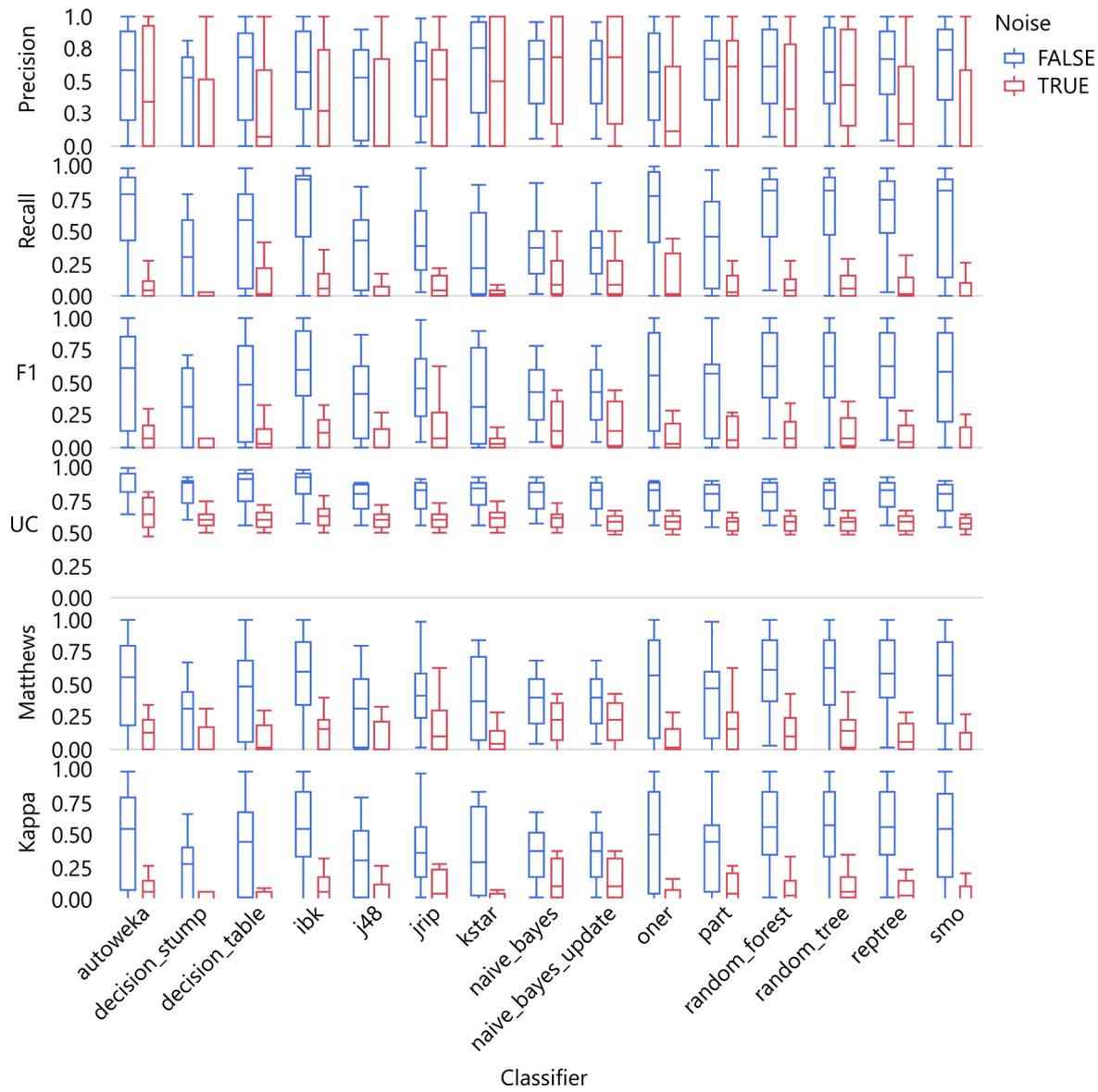

Fig. 5. Distribution of performance, among datasets, of different classifiers (x-axis), in the cases of with, or without, the snoring.

**RQ₁ Summary:** The effect of snoring on classification performance is of large size on all classifiers and all metrics other than Precision.

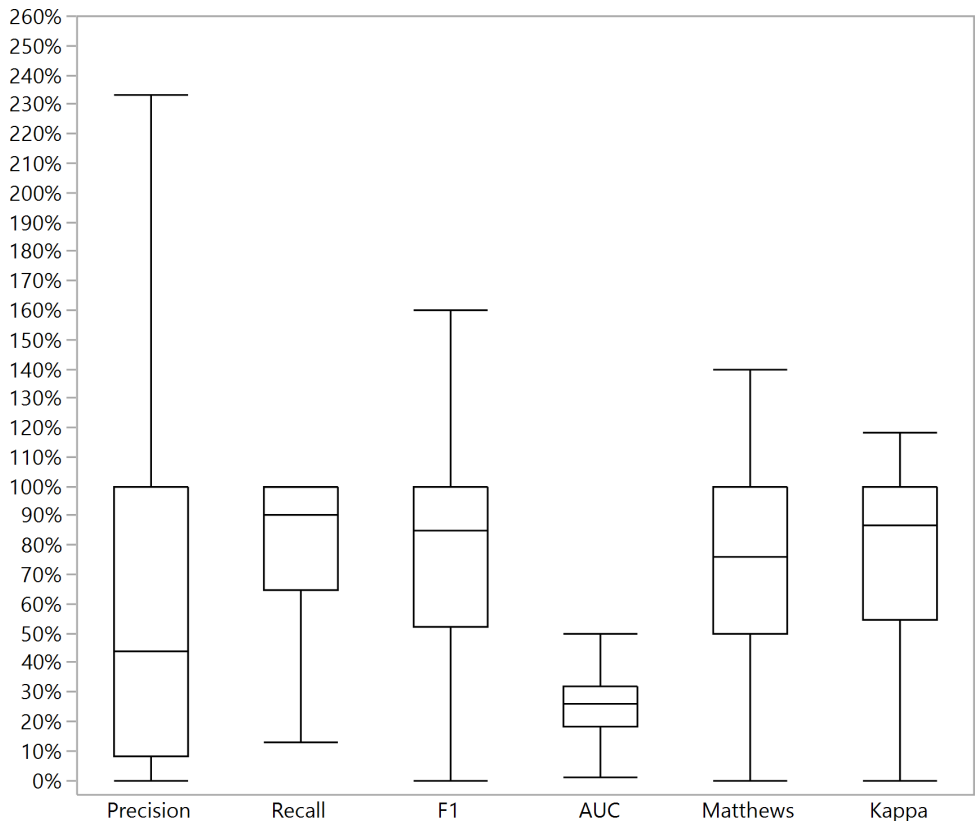

Fig. 6. Distribution of relative loss, among classifiers and datasets.

## 3.2 RQ₂: Does defect prediction performance improve if we remove from the training set the classes that are labeled as non-defective in the last releases?

Figure 7 reports the distribution among datasets and classifiers, of different performance metrics (x-axis), when removing a specific number of last releases data (color). Table 9 summarizes Figure 7 by reporting the relative gain (%), on average among datasets and classifiers, by removing a specific number of releases of data, on average across classifiers and datasets. According to Table 9:

(1) **Removing one release is better than removing no release in all performance metrics**.
(2) Removing three or four releases is worse than removing one or two releases in all performance metrics.
(3) Removing two releases is better than removing one release only in terms of Recall. In other words, removing one release is better than removing two releases in Precision and all the combined performance metrics.
(4) The gain in removing releases is particularly small in terms of AUC.
(5) Removing more than one release reduces Precision.

Figure 8 reports the distribution of performance in each classifier, among projects, by removing, or not, non-defective classes in the last release of the training set. According to Figure 8, removing one release is better than removing no release of data for all classifiers and all performance metrics.

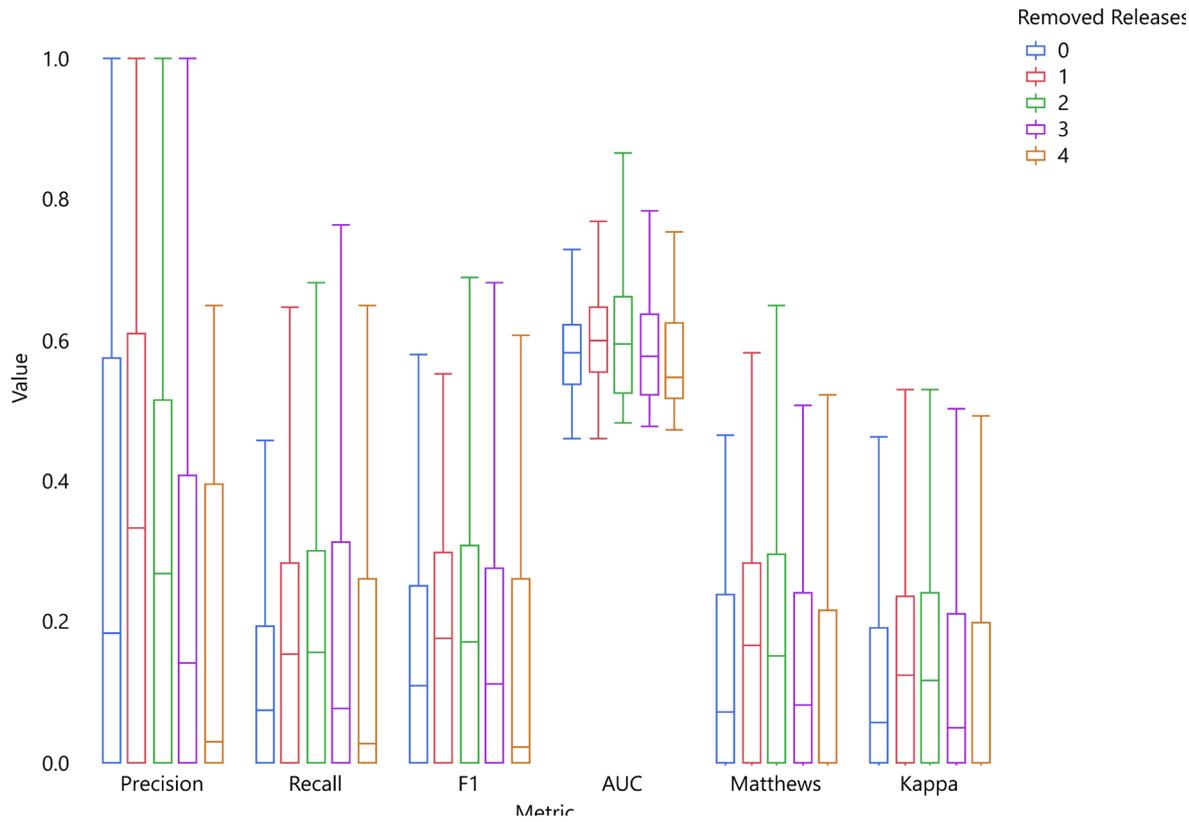

Fig. 7. Distribution of performance, among datasets, in the cases of removal of a specific number of last releases of data.

Table 9. Relative gain (%) in performance achieved by removing a specific number of releases of data, on average across classifiers and datasets.

| Removed Releases | Precision | Recall | F1 | Kappa | Matthew | AUC |
|---|---|---|---|---|---|---|
| 1 | 17% | 41% | 33% | 39% | 32% | 3% |
| 2 | -1% | 67% | 37% | 44% | 31% | 3% |
| 3 | -27% | 45% | 7% | -3% | -12% | 2% |
| 4 | -39% | 29% | 9% | -18% | -25% | 0% |

Table 10 reports the permutation test results on the difference in performance, in the cases of removing, or not, non-defective classes in the last release of the training set. According to Table 10, we can reject $H_20$ and claim that removing non-defective classes in the last release of the training set improves the performance of classifiers in all performance metrics other than Recall. Moreover, according to Table 10, the performance of classifiers varies across classifiers on all performance metrics.

Table 11 reports the Spearman correlation between the gain in performance, provided by removing non-defective classes in the last release of the training set, as measured by Precision and Recall, and the frequency of releases in projects, as computed in terms of the average number of days in a release and the average number of commits in a release. According to Table 11 there is a relevant and almost statistically significant inverse correlation between the gain in performance, provided by removing non-defective classes in the last release of the training set, and the number of commits per release. Thus, the smaller are the releases in terms of the number of commits, the higher is the gain in non-defective classes in the last release of the training set.

Table 10. RQ$_3$: Permutation test results on the effect of dropping releases on classifier performance metrics.

| | Precision | | | | |
|---|---|---|---|---|---|
| | Df | R Sum Sq | R Mean Sq | Iter | Pr(Prob) |
| **Between projects effects:** | | | | | |
| Classifier | 1 | 0.13 | 0.13 | 51.00 | 0.80 |
| Residuals | 10 | 12.41 | 1.24 | | |
| **Within project effects:** | | | | | |
| Classifier | 16 | 10.41 | 0.65 | 5000.00 | 0.00 |
| DropCount | 1 | 2.20 | 2.20 | 5000.00 | 0.00 |
| Classifier:DropCount | 16 | 1.79 | 0.11 | 5000.00 | 0.02 |
| Residuals | 945 | 55.88 | 0.06 | | |
| | Recall | | | | |
| | Df | R Sum Sq | R Mean Sq | Iter | Pr(Prob) |
| **Between projects effects:** | | | | | |
| Classifier | 1 | 0.00 | 0.00 | 51.00 | 0.92 |
| Residuals | 10 | 7.40 | 0.74 | | |
| **Within project effects:** | | | | | |
| | Df | R Sum Sq | R Mean Sq | Iter | Pr(Prob) |
| Classifier | 16 | 3.91 | 0.24 | 5000.00 | 0.00 |
| DropCount | 1 | 0.09 | 0.09 | 5000.00 | 0.01 |
| Classifier:DropCount | 16 | 1.02 | 0.06 | 5000.00 | 0.00 |
| Residuals | 945 | 25.51 | 0.03 | | |
| | Kappa | | | | |
| | Df | R Sum Sq | R Mean Sq | Iter | Pr(Prob) |
| **Between projects effects:** | | | | | |
| Classifier | 1 | 0.21 | 0.21 | 169.00 | 0.37 |
| Residuals | 10 | 4.11 | 0.41 | | |
| **Within project effects:** | | | | | |
| DropCount | 1 | 0.09 | 0.09 | 5000.00 | 0.00 |
| Classifier | 16 | 2.01 | 0.13 | 5000.00 | 0.00 |
| DropCount:Classifier | 16 | 0.36 | 0.02 | 5000.00 | 0.20 |
| Residuals | 945 | 15.05 | 0.02 | | |
| | MCC | | | | |
| | Df | R Sum Sq | R Mean Sq | Iter | Pr(Prob) |
| **Between projects effects:** | | | | | |
| Classifier | 1 | 0.11 | 0.11 | 51.00 | 0.76 |
| Residuals | 10 | 4.70 | 0.47 | | |
| **Within project effects:** | | | | | |
| DropCount | 1 | 0.22 | 0.22 | 5000.00 | 0.00 |
| Classifier | 16 | 2.69 | 0.17 | 5000.00 | 0.00 |
| DropCount:Classifier | 16 | 0.45 | 0.03 | 5000.00 | 0.03 |
| Residuals | 945 | 18.26 | 0.02 | | |
| | AUC | | | | |
| | Df | R Sum Sq | R Mean Sq | Iter | Pr(Prob) |
| **Between projects effects:** | | | | | |
| Classifier | 1 | 0.00 | 0.00 | 51.00 | 0.98 |
| Residuals | 10 | 2.59 | 0.26 | | |
| **Within project effects:** | | | | | |
| DropCount | 1 | 0.00 | 0.00 | 924.00 | 0.10 |
| Classifier | 16 | 0.67 | 0.04 | 5000.00 | 0.00 |
| DropCount:Classifier | 16 | 0.00 | 0.00 | 483.00 | 1.00 |
| Residuals | 945 | 3.05 | 0.00 | | |

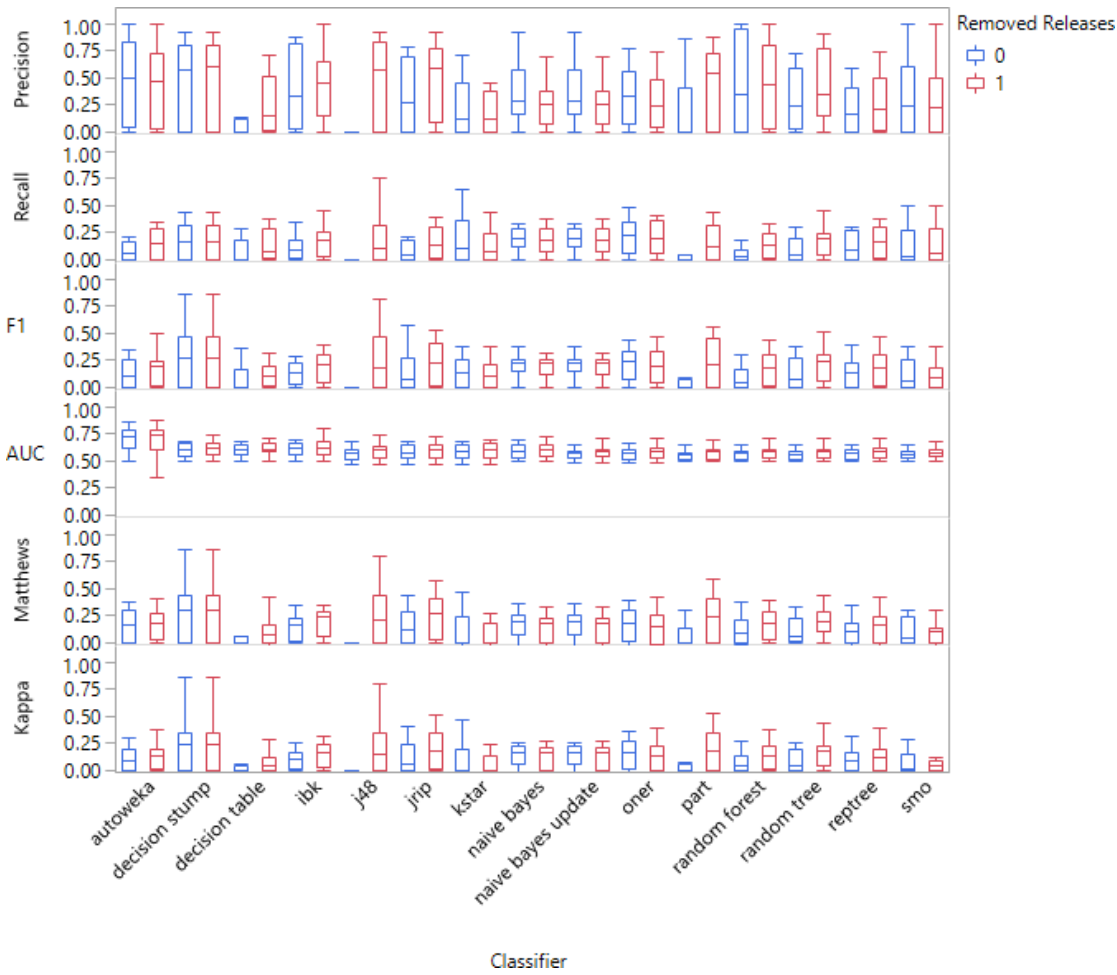

Fig. 8. Distribution of performance in each classifiers, among projects, by removing, or not, non-defective classes in the last release of the training set.

Table 11. Spearman correlation between the gain in performance, provided by removing non-defective classes in the last release of the training set, and the frequency of releases in projects.

| Frequency | Metric | Rho | Pvalue |
|---|---|---|---|
| Days per release | Precision | -0.0979 | 0.7621 |
| Commits per release | Precision | -0.5734 | 0.0513 |
| Days per release | Recall | 0.2098 | 0.5128 |
| Commits per release | Recall | 0.0839 | 0.7954 |

> **RQ$_2$ Summary:** Using a training set that excludes non-defective classes from the last one or two releases increases the classifiers' performance in all six metrics. However, further exclusion worsens the models' performance.

## 4 DISCUSSION

This section describes the implications of the existence of snoring, on the performances of machine learning models to predict class defectiveness.

Based on RQ$_1$ results, the presence of snoring significantly decreases the classifiers' performance. For instance, the median Recall achieved by training a classifier on the entire dataset with snoring is 90% less than the Recall ideally achievable by training that same classifier on the same entire dataset without snoring.

It is interesting to note that we found no significant differences in what concerns Precision. One possible reason for this high impact of snoring on Recall and low impact on Precision is the fact that snoring likely increases the proportion of negatives in the training set. Thus, training classifiers providing a high number of negatives, which result in a high proportion of false negatives and hence a lower Recall.

Unfortunately, the snoring instances cannot be accurately identified yet, and therefore, an entire dataset without snoring is only ideal. What we know is that, by definition, the snoring classes are more frequent in last releases than in early releases. However, regardless of the number of releases we ignore, we know that we also ignore data that is valid, i.e., classes that are actually non-defective. The impossibility to remove only noise instances implies a trade-off between the amount of data and the quality of the data used to feed prediction models.

Results from RQ$_2$ suggest that performances improve when removing non-defective data from the last release, and further improve on all metrics other than Precision when removing the last two releases. At the same time, going way too ahead, removing more than two releases, means training the model with less and older data, and then worsen the models' performances. Specifically, according to the RQ$_2$ results, removing the last two releases of data is better than removing only the last release in terms of Recall, but worse in terms of Precision. Therefore, based on RQ$_2$ results, a good trade-off is to not train classifiers with classes labeled as non-defective in the last release.

Researchers should identify which specific class in the last releases is erroneously labeled as non-defective. This advancement would allow a reduction in the number of removed non-defective classes and hence an improvement in the amount of data that can be used to feed prediction models.

## 5 THREATS TO VALIDITY

In this section, we discuss the study threats to validity, and specifically construct, internal, conclusion, and external validity threats [78].

*Construct validity* threats are related to the relationship between theory and observation. The first construct validity threat is related to the dataset construction, i.e., how we created a dataset in which we know where snoring occurs. As explained in Section 2.2.3, we achieved this goal by removing the last 50% of the dataset. Although previous work suggests that this may drop about 99% of the snoring [3], we cannot exclude the presence of some snoring that we did not capture. To identify buggy classes, we combined the availability of the JIRA "affected version" with the use of the SZZ algorithm. The former provides a more realistic assessment of buggy classes, however, it is not always available, and for this reason, we combined it with SZZ. However, we are aware that SZZ determines the affected versions differently than what happens with the manually-inserted label, i.e., considering all versions from the bug introduction until its fixing.

Table 12. Proportion of defective classes in the same training set when removing classes labeled as defective in the last 0,1,2,3, and 4 releases of the training set.

| Project Name | Defective - 0 | Defective - 1 | Defective - 2 | Defective - 3 | Defective - 4 |
|---|---|---|---|---|---|
| AVRO | 0.03 | 0.04 | 0.05 | 0.08 | 0.12 |
| BOOKKEEPER | 0.19 | 0.23 | 0.3 | 0.42 | 0.62 |
| CHUKWA | 0 | 0 | 0 | 0 | 0 |
| CONNECTORS | 0.02 | 0.03 | 0.03 | 0.04 | 0.07 |
| CRUNCH | 0.09 | 0.12 | 0.18 | 0.31 | 0.76 |
| FALCON | 0.01 | 0.01 | 0.02 | 1 | 1 |
| GIRAPH | 0.15 | 0.22 | 0.29 | 0.42 | 0.76 |
| IVY | 0.08 | 0.08 | 0.1 | 0.11 | 0.13 |
| OPENJPA | 0.04 | 0.05 | 0.06 | 0.1 | 0.16 |
| PROTON | 0.02 | 0.02 | 0.02 | 0.02 | 0.02 |
| SSHD | 0.1 | 0.13 | 0.16 | 0.21 | 0.3 |
| STORM | 0.09 | 0.11 | 0.12 | 0.14 | 0.17 |
| SYNCOPE | 1.16 | 1.16 | 1.32 | 1.54 | 1.67 |
| TAJO | 0.23 | 0.33 | 0.5 | 0.78 | 1.17 |
| TEZ | 0.08 | 0.18 | 1 | 1 | 1 |
| THRIFT | 0.14 | 0.17 | 0.21 | 0.27 | 0.37 |
| WHIRR | 0.04 | 0.06 | 0.1 | 0.25 | 1 |
| ZEPPELIN | 0.17 | 0.22 | 0.3 | 0.39 | 0.57 |
| ZOOKEEPER | 0.05 | 0.07 | 0.08 | 0.11 | 0.17 |

Another threat to construct validity is the absence of the use of any balancing technique. The impact of balancing on classifiers' accuracy has been analyzed in a context similar to ours such as just-in-time classifiers (JIT) [68]. As stressed for JIT, and known in other domains [77], balancing techniques generally improve Recall at the cost of reducing Precision with no impact on overall accuracy performance metrics. Table 12 reports the proportion of defective classes in the same training set when removing classes labeled as defective in the last [0-4] releases of the training set. According to Table 12, removing non-defective data from the last release resulted in a slightly more balanced dataset. Since the observed improvement is in terms of Precision, Recall, and AUC, then we are confident that the observed increment is due to less noise in the data rather than a more balanced dataset. It was surprising to observe an increment in Precision despite a more balanced dataset; i.e., the snoring-free data counterbalanced the negative effect on Precision of a more balanced dataset.

*Internal validity* threats concern variables internal to our study and not considered in our experiment that could influence our observations on the dependent variable. In our study, we split data into a 66% training set and a 33% test set. However, it is possible that different splits could produce varying results. As our results already show, the effect of dormant defects can vary based on the choice of a specific machine learning model used for defect prediction. We cannot exclude that other models not considered in our study could exhibit different performance variations when the dataset is affected by dormant defects and when not. Moreover, it is also possible that tuning the models' hyperparameters would produce different results as well. However, it is not our goal to find the best performing machine learning model for defect prediction, but to simply show that dormant defects have a significant effect on the models' performances. The paper shows this is the case on all accuracy metrics other than recall.

*Conclusion validity* threats regard issues that affect the ability to draw accurate conclusions about relations between the treatments and the outcome of an experiment [78]. We see no major threat to conclusion validity.

*External validity threats* concern the generalizability of our results, and are especially related to our choice and the representativeness of the artifacts (projects) considered in the study. This study used 19 datasets and hence could be deemed of low generalization compared to studies using hundreds of datasets. However, our datasets are large if we considered the number of releases (616) and defects (4101) and as stated by Nagappan et al. [51], "*more is not necessarily better.*". Finally, we preferred to test our hypotheses only on datasets in which we were confident quality is high and that are close to industry.

## 6  RELATED WORK & BACKGROUND

In recent and past years, several studies investigated approaches for creating defect prediction datasets [12, 16, 39, 53, 62, 66] such as SZZ and others. Other studies investigated how to select these datasets [18, 22, 51, 63].

Several authors have discussed factors hindering the accuracy and evaluation of defect prediction models [1, 5, 20, 46, 47, 70, 72, 80]. With respect to the aforementioned work, we empirically show that snoring constitutes an additional factor hindering their accuracy.

Chen et al. [8] firstly introduced the concept of *dormant defects* (though called dormant bugs). They showed that dormant defects are fixed faster and by more experienced developers. Similarly, Rodriguez-Perez et al. [61] and da Costa et al. [13] show that the time to fix a defect, i.e. the sleeping phenomenon, is on average about one year. Thus, we conclude that dataset creation will miss most defects on releases that are less than a year old. In a recent paper [3] we reported that many defects are dormant and many classes snore. Specifically, on average among projects, most of the defects in a project are dormant for more than 20% of the existing releases, and 2) in the majority of the projects the missing rate is more than 25% even if we remove the last 50% of releases. Concerning previous work on dormant defects, including our, in this paper, we quantify, for the first time, the effect of dormant defects on defect prediction accuracy, their evaluation, and we provide and evaluate a countermeasure.

A problem of paramount importance when creating defect prediction models is represented by the presence of noise in the underlying datasets. In this context, different works study the sources of noise and foresee countermeasures for that. Kim et al. [36] measured the impact of noise on defect prediction models and provided guidelines for acceptable noise levels. They also propose a noise detection and elimination algorithm to address this problem. However, the noise studied and removed is supposed to be random. Tantithamthavorn et al. [69] found that: (1) issue report mislabelling is not random; (2) precision is rarely impacted by mislabeled issue reports, suggesting that practitioners can rely on the accuracy of modules labeled as defective by models that are trained using noisy data; (3) however, models trained on noisy data typically achieve about 60% of the recall of models trained on clean data. Herzig et al. [27] reported that 39% of files marked as defective never had a defect actually. They discuss the impact of this misclassification on earlier studies and recommend manual data validation for future studies. Rahman et al. [60] showed that size always matters as much as bias direction, and in fact, much more than bias direction when considering information-retrieval measures such as AUCROC and F-score. This result indicates that at least for prediction models, even when dealing with sampling bias, simply finding larger samples can sometimes be sufficient. Bird et al. [6] found that bias is a critical problem that threatens both the effectiveness of processes that rely on biased datasets to build prediction models and the generalizability of hypotheses tested on biased data. Similarly to previous work, we study a source of noise in defect prediction datasets, i.e., snoring.

McIntosh and Kamei [45] recently found that in the context of Just In Time (JIT) prediction, training prediction models on the last three months of data provides more accurate results than training on more past data. Their result seem to be in contrast to ours; we observed more accurate results by ignoring recent data; they observed more accurate results by ignoring old data. There are several possible reasons for this inconsistency. First, the different contexts; we classify classes whereas they classify commit. Second the analyzed projects: they analyze two projects that are different from our 19 ones. Third, the two results might be consistent; it could be that the optimal data to train the models is the one in the middle, i.e., not too old (as showed in McIntosh and Kamei [45]), not too recent (as showed here). Thus, data for prediction models might act as vegetables, and you do not want them either unripe or rotten.

## 7 CONCLUSION

Dormant defects refer to defects discovered several releases after their introduction [8, 61]. While this concept is well-known in the software engineering community, its impact on defect prediction models was not investigated before. In this paper, we empirically analyze the effect of dormant defects in causing classes to be tagged as non-defective in defect prediction dataset and therefore affect the performance of defect prediction models. We analyze data from more than 4,000 defects and 600 releases of 19 open source projects from the Apache ecosystem and 15 machine learning defect prediction classifiers. Our results show that (i) the presence of snoring decreases the recall of defect prediction classifiers, while not significantly affecting precision; and (ii) removing data from recent releases significantly improves the classifiers' performance. At the same time, going too behind worsen performances. In summary, this paper provides insights on how to create a software defect dataset by mitigating the effect of snoring on classifiers performances.

While, in principle, using recent data could be valuable when building a defect prediction dataset (i.e., the system characteristics are more similar to the current status), this paper showed how recent releases may contain defects that have not been discovered yet. For such a reason, classes with only such defects become "dormant" and represent false cases of defect-free classes in the dataset. Without doing better analysis and testing there is no simple way to discover defects in recent versions. At the same time, this paper empirically shows the extent to which relying on less recent releases reduces the snoring effect in defect prediction results.

In terms of future work, we plan to extend this research by:

(1) *Combining noise removal techniques:* There are several types of noise currently known in defects datasets, including snoring and misclassification [27, 69]. However, no approach can remove all noises from a dataset.
(2) *Considering distribution as inputs for the defect prediction models*: We envision a model that accepts, as an input, a distribution of values, rather than a point-value [64]. In other words, since we cannot be sure about the defectiveness of a class in the training set, then, intuitively, the defectiveness of a class should be modelled by a confidence interval rather than a binary measure.
(3) *Analyzing the impact of snoring on existing research on factors affecting the performance of defect prediction:* previous research has shown the impact of tuning [1, 20] or balancing [68] on defect prediction. Such an impact could be influenced by snoring, and it would be therefore interesting to investigate that effect.
(4) *Application in the context of Just in Time Defect Prediction (JIT):* JIT prediction models have become sufficiently robust so that they are now incorporated into the development cycle of some companies [45]. We can replicate the present study on JIT models to understand the effect snoring has on the performance of these models.